\documentclass{sig-alternate-05-2015}
\usepackage{bbm}
\usepackage{multirow}
\usepackage{gensymb}
\usepackage{mathtools}
\usepackage{hyperref}
\usepackage{subfig}
\usepackage{floatrow}

\begin{document}


\newcommand{\Pbb}{\mathbb{P}}

\newcommand{\Xc}{\mathcal{X}}
\newcommand{\Yc}{\mathcal{Y}}
\newcommand{\Vc}{\mathcal{V}}

\def\eqd{\,{\buildrel d \over =}\,}

\newtheorem{theorem}{Theorem}
\newtheorem{proposition}{Proposition}
\newdef{definition}{Definition}
\newdef{assumption}{Assumption}
\newtheorem{remark}{Remark}




\acmPrice{\$15.00}

%

\title{Privacy-Enhanced Architecture for Occupancy-based HVAC Control}

%
%
%
%
%
\numberofauthors{1} 
%
\author{
%
%
\alignauthor
Ruoxi Jia\textsuperscript{1}, Roy Dong\textsuperscript{2}, S. Shankar Sastry\textsuperscript{2}, Costas J. Spanos\textsuperscript{1}\\
       \affaddr{Department of Electrical Engineering and Computer Sciences}\\
       \affaddr{University of California, Berkeley}\\
       \email{ruoxijia@berkeley.edu,roydong@eecs.berkeley.edu}\\
       \email{sastry@eecs.berkeley.edu,spanos@berkeley.edu}
}


\maketitle
\begin{abstract}

Large-scale sensing and actuation infrastructures have allowed buildings to achieve significant energy savings; at the same time, these technologies introduce significant privacy risks that must be addressed. In this paper, we present a framework for modeling the trade-off between improved control performance and increased privacy risks due to occupancy sensing. More specifically, we consider occupancy-based HVAC control as the control objective and the location traces of individual occupants as the private variables. Previous studies have shown that individual location information can be inferred from occupancy measurements. To ensure privacy, we design an architecture that distorts the occupancy data in order to hide individual occupant location information while maintaining HVAC performance. Using \emph{mutual information} between the individual's location trace and the reported occupancy measurement as a privacy metric, we are able to optimally design a scheme to minimize privacy risk subject to a control performance guarantee. We evaluate our framework using real-world occupancy data: first, we verify that our privacy metric accurately assesses the adversary's ability to infer private variables from the distorted sensor measurements; then, we show that control performance is maintained through simulations of building operations using these distorted occupancy readings. 

\end{abstract}
\footnotetext[1]{This research is funded by the Republic of Singapore's National Research Foundation through a grant to the Berkeley Education Alliance for Research in Singapore (BEARS) for the Singapore-Berkeley Building Efficiency and Sustainability in the Tropics (SinBerBEST) Program. BEARS has been established by the University of California, Berkeley as a center for intellectual excellence in research and education in Singapore.}
\footnotetext[2]{This research is supported in part by FORCES (Foundations Of Resilient CybEr-Physical Systems), which receives support from the National Science Foundation (NSF award numbers CNS-1238959, CNS-1238962, CNS-1239054, CNS-1239166).}

%
%
\newcommand\HH{
  \global\let\savedtextbullet\textbullet
  \gdef\textbullet{%
    \par\noindent\savedtextbullet\global\let\textbullet\savedtextbullet
  }%
}


%
%

%
%



\section{Introduction}

Large-scale sensing and actuation infrastructures have endowed buildings with the intelligence to perceive the status of their environment, energy usage, and occupancy, and to provide fine-grained and responsive controls over heating, cooling, illumination, and other facilities. However, the information that is collected and harnessed to enable such levels of intelligence may potentially be used for undesirable purposes, thereby raising the question of privacy. To spotlight the value of building sensory data and its potential for exploitation in the inference of private information, we consider as a motivating example the {occupancy} data, i.e., the number of occupants in a given space over time.

Occupancy data is a key component to perform energy-efficient and user-friendly building management. Particularly, it offers considerable potential for improving energy efficiency of the heating, ventilation, and air conditioning (HVAC) system, a significant source of energy consumption which contributes to more than $50\%$ of the energy consumed in buildings~\cite{eia2011annual}. Recent papers~\cite{balaji2013sentinel,kleiminger2014smart,erickson2010occupancy} have demonstrated substantial energy savings of up to $40\%$ by enabling intelligent HVAC control in response to occupancy variations. The value of occupancy data in building management has also inspired extensive research on occupancy sensing~\cite{dong2010information,jin2015sensing,jin2014presencesense,khan2015infrastructure,yang2015cross} as well as a number of commercial products which can provide high accuracy occupancy data. 

\begin{figure}[!t]
\centering
\includegraphics[width=\columnwidth,trim={0cm 0.02cm 0cm 4.18cm},clip]{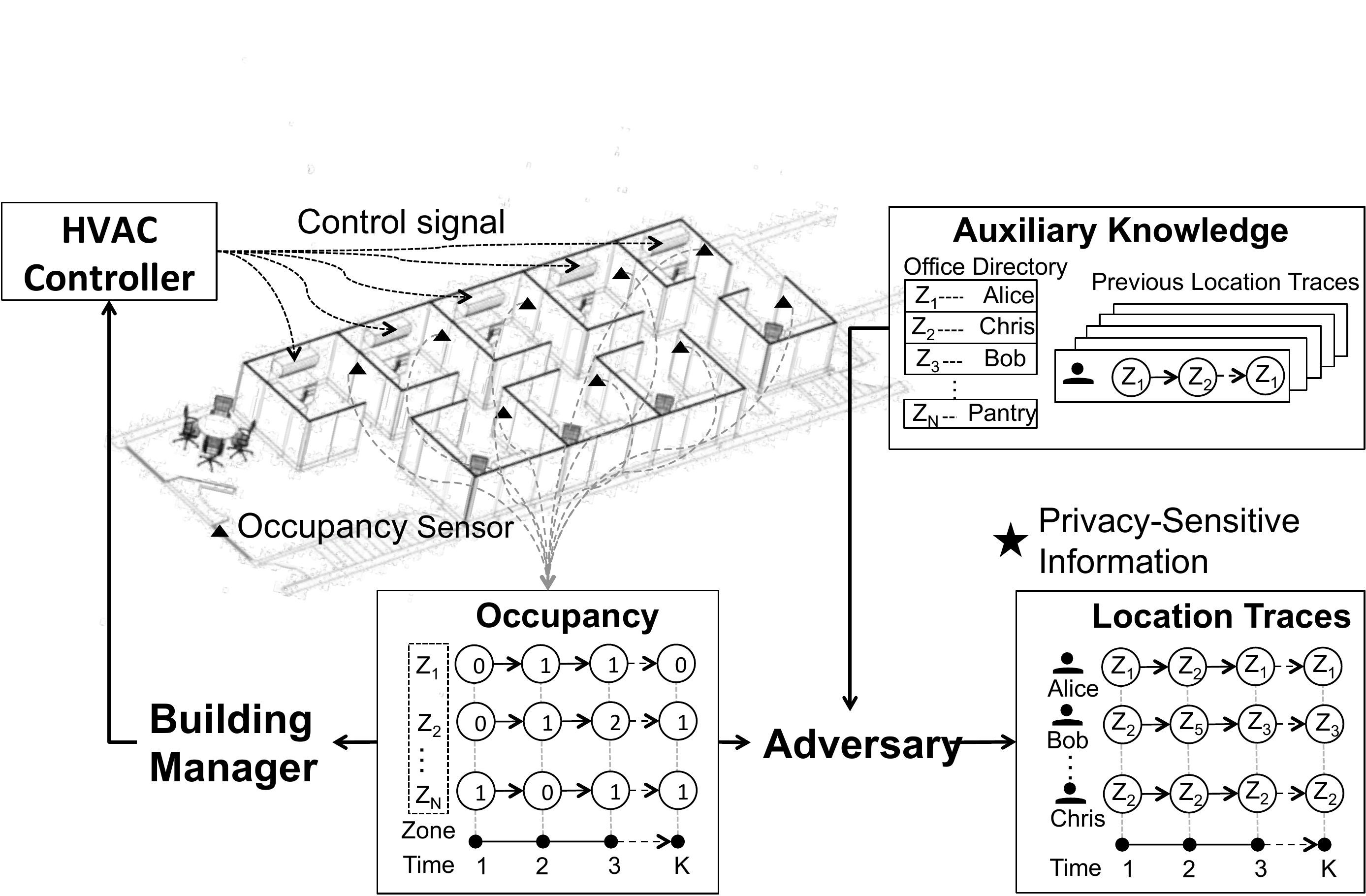}
\caption{An overview of the problem of individual occupant location recovery. The building manager collects occupancy data to enable intelligent HVAC controls adapted to occupancy variations. However, an adversary with malicious intent may exploit occupancy data in combination with the auxiliary information to infer privacy details about indoor locations of building users.}
\label{fig:prob_overview}
\end{figure}

While people have enjoyed the benefits brought by occupancy data, the privacy risks potentially posed by the data are largely overlooked (Figure~\ref{fig:prob_overview}). In effect, location traces of individual occupants can be inferred from the occupancy data with some auxiliary information~\cite{wang2014non}. Throughout this paper, we refer to the individual location trace as the private information to be protected. The contextual information attached to location traces tells much about the individuals' habits, interests, activities, and relationships~\cite{Lisovich2010}. It can also reveal their personal or corporate secrets, expose them to unwanted advertisement and location-based spams/scams, cause social reputation or economic damage, make them victims of blackmail or even physical violence~\cite{shokri2011quantifying}. 

At a first glance, it is surprising that occupancy data may incur risks of privacy breach, since it only reports the number of occupants in a given space over time without revealing the identities of the occupants. To illustrate why it is possible to infer location traces from seemingly ``anonymized" occupancy data, consider the following scenario. We start by observing two users in one room and then one of them leaves the room and enters another room. We cannot tell which one of the two made this transition by observing the occupancy change. However, if the one who left entered an private office, the user can be identified with high probability based on the ownership of the office. Although a change in occupancy data may correspond to location shifts of many possible potential users, the knowledge of where the individuals mostly spend their time rules out many possibilities and renders the individual who made the transition identifiable. It has been shown in~\cite{wang2014non} that by simply combining some ancillary information, such as an office directory and user mobility patterns, individual location traces can be inferred from the occupancy data with the accuracy of more than $90\%$. It is, therefore, the objective of this paper to enable an occupancy-based HVAC control system that provides privacy features for each user on a par with thermal comfort and energy efficiency.

A simple yet effective way to preserve privacy is to obfuscate occupancy data by injecting noise to make the data itself less informative. This approach has been widely used in privacy disclosure control of various databases, ranging from healthcare~\cite{dankar2013practicing}, geolocation~\cite{andres2013geo}, web-browsing behavior data~\cite{fan2014monitoring}, etc. While reducing the risk of privacy breach, this approach would also deteriorate the utility of the data. There have been attempts to balance learning the statistics of interest reliably with safeguarding the private information~\cite{soria2014enhancing}. Cryptography~\cite{diffie1979privacy} and access control~\cite{WangSunBertino2014} are also effective means to ease privacy concerns, but they do not provide protection against all privacy breaches. There may be insiders who can access the private, decrypted data, or the building manager may not want to have access to (and responsibility for) the private data.



The objective of this paper cannot be attained by simply extending the techniques developed previously. Our task is more challenging. Firstly, as opposed to learning some fixed statistics from static data in most database applications, the data is used for controlling a highly complex and dynamic system in our case, and the control performance relies on the data fidelity. With highly accurate occupancy data, the infrastructure can correctly sense the environment and enable proper response to occupancy variations; nevertheless, the location privacy is sacrificed. On the other hand, the usage of severely distorted occupancy data reduces the risks of privacy leakage, but may lead to even higher levels of energy consumption and discomfort. Essentially, we need to address the trade-off between the performance of a controller on a dynamical system, and, similarly, privacy of a time-varying signal, i.e. the location traces of individual occupants. Secondly, from the perspective of the building manager, the building performance is paramount: adding the privacy feature into the HVAC control system should not impair the performance of HVAC controller in terms of energy efficiency and thermal comfort. 
To achieve this, the injected noise should be calculated to minimally affect performance of the controller, while maximizing the amount privacy gained from the distortion.






In this paper we develop a method which minimizes the privacy risks incurred by collection of occupancy data while guaranteeing the HVAC system operating in a ``nearly" optimal condition. Our solution relies on an occupancy distortion mechanism, which informs the building manager how to distort occupancy data before any form of storage or transport of the data. We draw the inspiration from the information-theoretic approach in~\cite{RajagopalanSankarMohajerEtAl2011,PinCalmon2012} for characterizing the privacy-utility trade-off, and choose the mutual information (MI) between reported occupancy measurements and individual location traces as our privacy metric. The design problem of finding the optimal occupancy distortion mechanism is cast as an optimization problem where the privacy risk is minimized for a set of constraints on the controller performance. This allows us to find points on the Pareto frontier in the utility-privacy trade-off, and to further analyze the economic side of privacy concerns~\cite{Ratliff2015}. The formulation can be easily generalized to 
resolve the tension between privacy and data utility in other cases where a control system utilizes some privacy-sensitive information as one of the control inputs, although in this paper we limit our focus to addressing the privacy concern of occupancy-based HVAC controller. In addition, our work here is complementary to the work being done in the cryptography communities: we can use our distortion mechanism to process sensor measurements, and then transmit the processed measurements across secure channels. Our work also serves as a complement for the privacy-preserving access control protocol in~\cite{WangSunBertino2014}, as it provides distortion mechanisms against adversaries who might be able to subvert the protocol while still retaining the benefits for the occupancy data. 

The main contributions of our paper are as follow:
\begin{itemize}
\item We present a systematic methodology to characterize the privacy loss and control performance loss.
\item We develop a holistic and tractable framework to balance the privacy pursuit and control performance.
\item We evaluate the trade-off between privacy and HVAC control performance using the real-world occupancy data and simulated building dynamics.
\end{itemize}

The rest of the paper is organized as follows: Section~\ref{sec:back} reviews the existing work on occupancy-based control algorithms and privacy metrics. Section~\ref{sec:model} describes the models connecting location and occupancy, and the HVAC system model that will be considered in this paper. In Section~\ref{sec:framework} we present a framework for quantifying the trade-off between privacy and controller performance. We will evaluate the framework and demonstrate its practical values based on experimental studies in Section~\ref{sec:eval}. Section~\ref{sec:conclusion} concludes the paper.

\section{Related Work}
\label{sec:back}

\subsection{Occupancy-based HVAC control}
Occupancy-based HVAC systems exploit real-time occupancy measurements to condition the space appropriate to usage. 
The occupancy-based controllers in the existing work can be categorized into two types: rule-based controller and optimization-based controller or model predictive control (MPC). The rule-based controller uses an ``if condition then action" logic for decision making in accordance with occupancy variations~\cite{erickson2010occupancy,balaji2013sentinel}. MPC is a more advanced control scheme, which employs a model of building thermal dynamics in order to predict the future evolution of the system, and solves an optimization problem in real-time to determine control actions~\cite{oldewurtel2012use}. A number of papers including~\cite{gyalistras2010use,hu2014model,aswani2012reducing} analyzed in large-scale simulative or experimental studies the energy saving potential in building climate control by using MPC, which was shown to be well-suited for building applications.
This leads to our choice of MPC to exemplify the trade-off between controller performance and privacy.

Occupancy information can be leveraged in different ways in an MPC-based controller. One approach is to build an occupancy model to predict future occupancy based on which the MPC optimizes control actions~\cite{beltran2014optimal}. Another method is to use the instantaneous occupancy measurement and consider it to be constant during the control horizon of MPC~\cite{goyal2013occupancy}. This method has been demonstrated to achieve comparable performance with the MPC that exploits occupancy predictions. We will thus without loss of generality follow the latter set-up to avoid explicit modeling of occupancy.

\subsection{Privacy}
\label{sec:back_privacy}
Privacy, although not a new topic, has recently developed renewed interest, due in no small part to new technologies and modern infrastructures collecting and storing unprecedented amounts of data. Since privacy is an abstract and subjective concept, it is necessary to develop proper measures for privacy before any privacy protection technique is discussed. 

Differential privacy~\cite{Dwork2006} is one of the most popular metrics for privacy from the area of statistical databases. It is is typically assured by adding appropriately chosen random noise to the database output. However, calculating optimal noise for differential privacy is very difficult, and
research on the applications of differential privacy mostly assumes the injected noise to be an additive zero-mean Gaussian or Laplacian random variable, which offers no guarantee on data utility. As mentioned in the introduction, in our case the performance of HVAC control systems is crucial: as such, our work is an effort to maintain control efficacy by optimally designing noise distribution to maximize privacy subject to a performance guarantee.

Recently, MI has become a popular privacy metric~\cite{RajagopalanSankarMohajerEtAl2011,PinCalmon2012,JiaoCourtadeVenkatEtAl2015}. Intuitively, MI reflects the change in the uncertainty of a private variable due to the observation of a public random variable. In fact, it is the \emph{only} metric of information leakage that satisfies the data processing inequality~\cite{JiaoCourtadeVenkatEtAl2015}. 
Unlike differential privacy, this requires some modeling of the adversary's available ancillary information; however, in practice, we can suppose an adversary with access to a large amount of ancillary information, which gives a bound on any weaker adversary's performance.
A framework for characterizing privacy-utility trade-off based on MI was proposed in~\cite{PinCalmon2012},
where the MI between a private variable and a distorted measurement is minimized subject to the bound on the value of an exogenous distortion metric that measures the utility loss from replacing a true measurement with a distorted measurement.
Our work is an extension of~\cite{PinCalmon2012} to the situations where dynamics at present. We propose a method to abstract out control performance of a dynamical system into a distortion metric, as well as a set of reasonable assumptions for the probabilistic dependencies between occupancy and location data, which allow us to re-write our privacy metric on time-series data into a static situation akin to that developed in~\cite{PinCalmon2012}.

\section{Preliminaries}
\label{sec:model}

This section collects the concepts we need before introducing the theoretical framework that characterizes the trade-off between privacy and control performance in Section~\ref{sec:framework}. Two models are described: the \textit{occupancy-location model} that formulates the relationship between occupancy observations and individual location traces, and the model for the HVAC system. We will first consider an occupancy detection system that can collect noise-free or true occupancy, which is then processed by a distortion mechanism into the obfuscated data that the controller observes. We will see the distortion can be similarly applied to noisy occupancy, as elaborated in Section~\ref{sec:framework}.

\subsection{Occupancy-location model}

Suppose the building of interest consists of $N$ zones represented by $\mathcal{Z} = \{z_0, z_1,\cdots,z_{N}\}$, where a special zone $z_{0}$ is added to refer to the outside of the building. Let $\mathcal{O}= \{o_1,\cdots,o_M\}$ denote the set of occupants. The location of occupant $o_m$ at time $k$ is a random variable denoted by $X_k^{(m)}$ which takes values in the set $\mathcal{Z}$, for $m = 1,\cdots,M$. The true occupancy of zone $z_n$ at time $k$ is denoted by $Y_k^n$, $n = 0,1,\cdots,N$. $Y_k^n$ takes values from $\{0,1,\cdots,M\}$, where $M$ is the total number of occupants in the building. Note that the true occupancy and individual location traces are connected by $Y_k^n = \sum_{m=1}^M \mathbbm{1}[X_k^{(m)}=z_n]$, where $\mathbbm{1}[\cdot]$ is the indicator function. 

Additionally, we suppose that the controller observes a distorted version of the true occupancy, denoted by $V_k^n$ which takes values from $\{0,1,\dots,M\}$. 
$\mathbb{P}(V_k^n|Y_k^n)$ represents the distortion mechanism we wish to design. 
If no distortion on the occupancy data is applied, then $V_k^n = Y_k^n$. We further define some shorthands: $X_k^{(1:M)}:=\{X_k^{(1)},\cdots,X_k^{(M)}\}$, $V_k^{1:N}:=\{V_k^1,\cdots,$ $V_k^N\}$.

We make the following assumptions.
\begin{assumption}
\label{ass:loc_indep}
The location traces for different occupants are mutually independent: $\Pbb(X_k^{(1:M)}) = \prod_{m = 1}^M \Pbb(X_k^{(m)})$.
\end{assumption}

\begin{assumption}
\label{ass:loc_markov}
The location trace for any given occupant $o_m$, $m \in \{1,\dots,M\}$, has the first-order Markov property:
\begin{equation}
\Pbb(X_k^{(m)} \vert X_{k-1}^{(m)},X_{k-2}^{(m)},\dots,X_{1}^{(m)}) =
\Pbb(X_k^{(m)} \vert X_{k-1}^{(m)})
\end{equation}
\end{assumption}

\begin{assumption}
\label{ass:loc_cond_indep}
The true occupancy $Y_k^n$ is a \emph{sufficient stat\-istics} for $V_k^n$, i.e., $\Pbb(V_k^n|X_k^{(1:M)}) = \Pbb(V_k^n|Y_k^n)$.
\end{assumption}

\begin{figure}[ht]
\centering
\includegraphics[width=0.9\columnwidth,trim={0cm 0.1cm 0cm 0cm},clip]{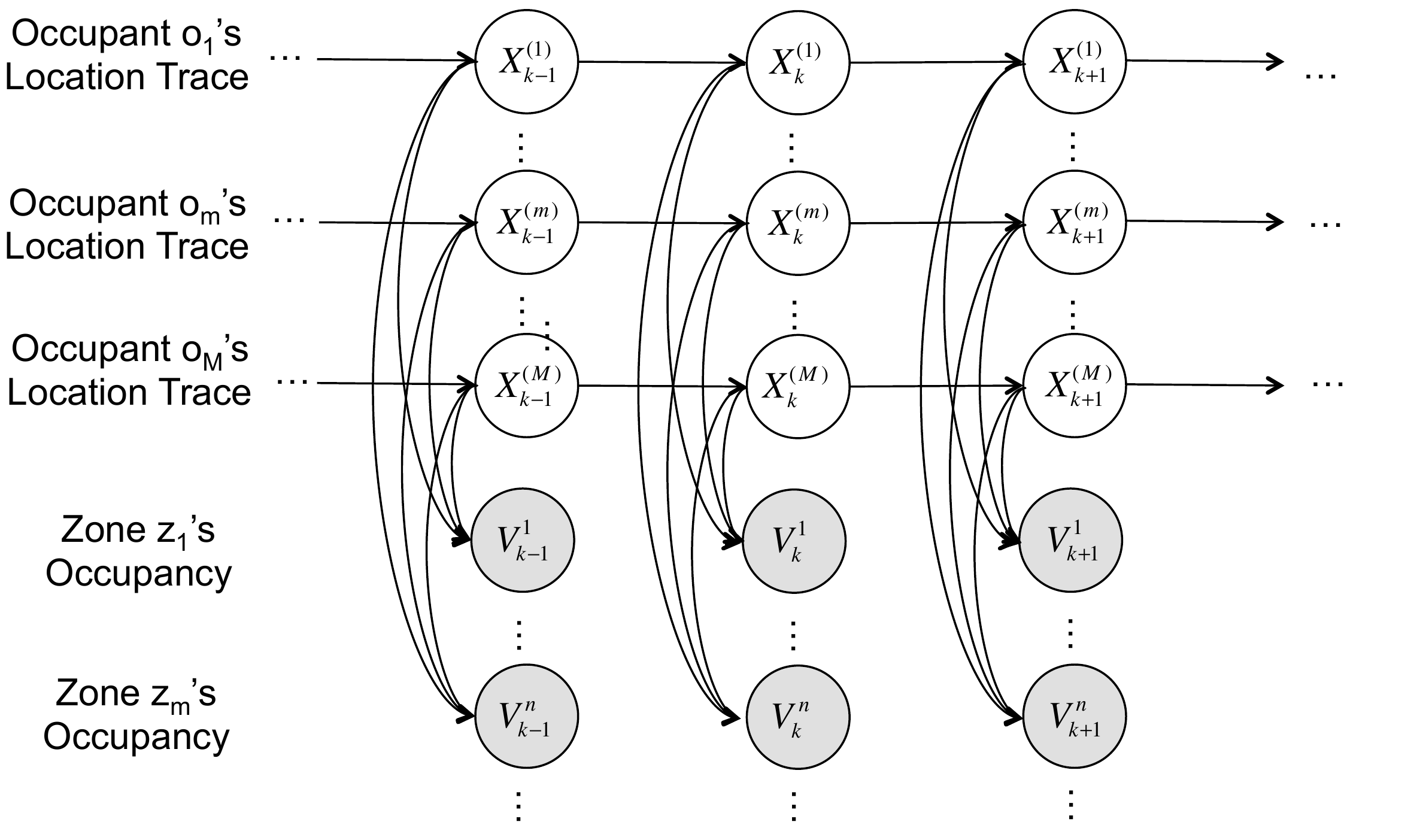}
\caption{The graphical model representation of the FHMM model.}
\label{fig:FHMM}
\end{figure}

Assumption~\ref{ass:loc_cond_indep} is naturally justified since the distribution of $V^n_k$ depends only on the value of $Y^n_k$ in our distortion mechanism. The first two assumptions are necessary to design the optimal distortion method, but we will show that our distortion method will work on the real-world occupancy dataset, which provides a support for Assumption~\ref{ass:loc_indep} and \ref{ass:loc_markov}. These assumptions allow us to model occupancy and location traces via the Factorial Hidden Markov model (FHMM), illustrated in Figure~\ref{fig:FHMM}. The FHMM consists of several independent Markov chains evolving in parallel, representing the location trace of each occupant. Since we only observe the aggregate occupancy information, the location traces are considered to be hidden states. 

The FHMM model can be specified by the transition probabilities and emission probabilities. The transition probabilities describe the mobility pattern of an occupant, which is denoted as a $(N+1)\times(N+1)$ transition matrix. We define the transition matrix for occupant $o_m$ as $A^{(m)}=[a_{ij}^{(m)}]$, $i,j=0,1,\cdots,N$, where $a_{ij}^{(m)}=\mathbb{P}(X_{k+1}^{(m)}=z_j|X_{k}^{(m)}=z_i)$ for $k=0,1,\cdots,K-1$. The transition parameters can be learned from the occupancy data based on maximum likelihood estimation. If the prior knowledge about the past location traces is also available, it can be encoded as the prior distribution of transition parameters from a Bayesian point of view, and then the transition parameters can be learned via \emph{maximum a posteriori} (MAP) estimation. We refer the readers to \cite{wang2014non} for the details of parameter learning. The emission probabilities characterize the conditional distribution of distorted occupancy given the location of each occupant, defined by
\begin{align}
\mathbb{P}(V_k^{1:N}|X_k^{(1:M)}) = \prod_{n=1}^{N} \mathbb{P}(V_k^n|X_k^{(1:M)})=\prod_{n=1}^{N}\mathbb{P}(V_k^n|Y_k^n)
\end{align}
The above equalities result from Assumption~\ref{ass:loc_cond_indep}, which, in other words, indicates that the distorted occupancy depends on individual location traces only via the true occupancy. 

\subsection{HVAC system model}
\label{sec:hvac_mode}
\begin{figure}[h]
\centering
\includegraphics[width=0.8\columnwidth,trim={0cm 0.3cm 0cm 0.3cm},clip]{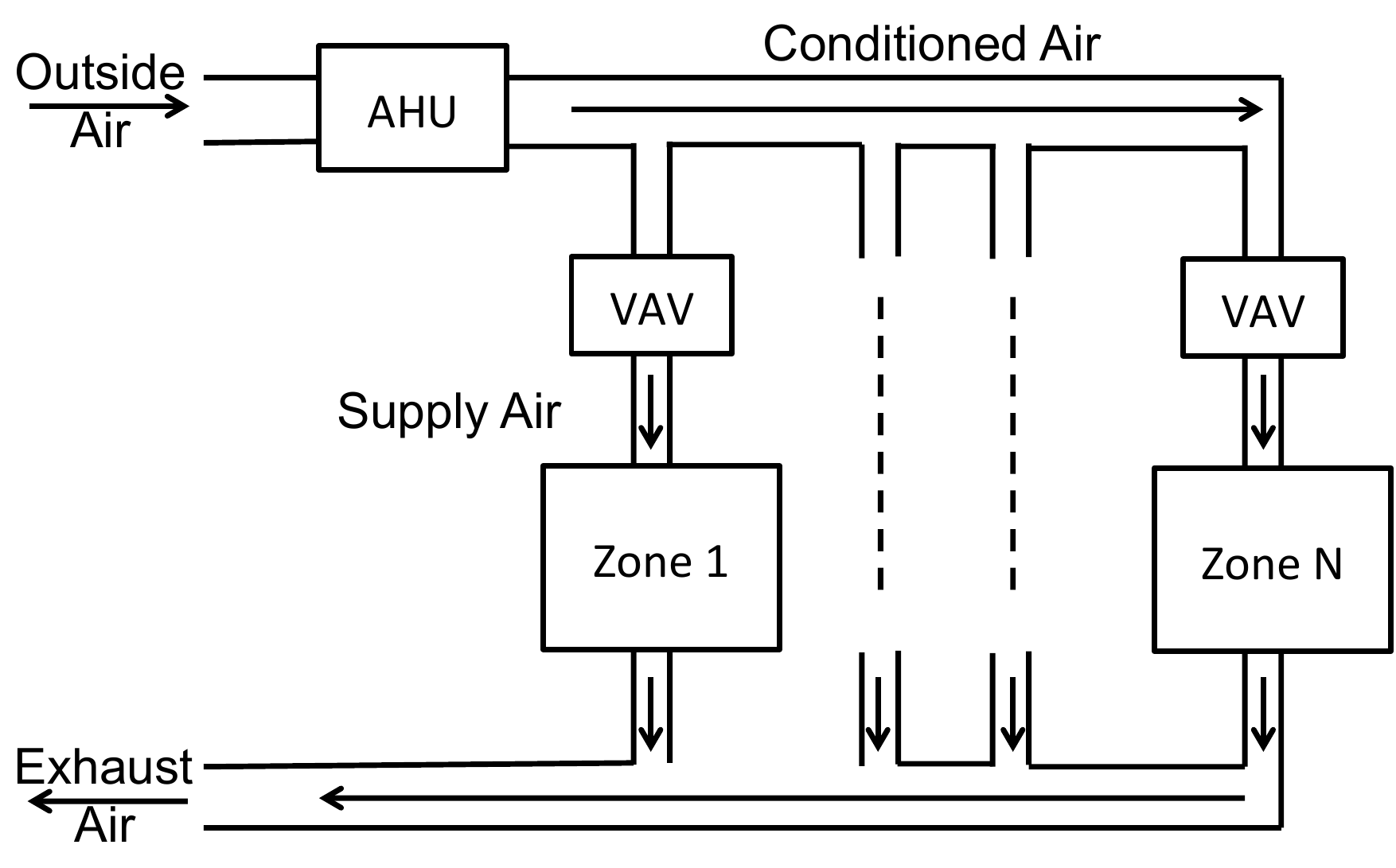}
\caption{A schematic of a typical multi-zone commercial building with a VAV-based HVAC system.}
\label{fig:hvac_scheme}
\end{figure}

Suppose the thermal comfort of the building space of interest is regulated by the HVAC system shown in Figure~\ref{fig:hvac_scheme}, which provides a system-wide Air Handling Unit (AHU) and Variable Air Volume (VAV) boxes distributed at the zones. In this type of HVAC system, the outside air is conditioned at the AHU to a setpoint temperature $T_a$ by the cooling coil inside. The conditioned air, which is usually cold, is then supplied to all zones via the VAV box at each zone. The VAV box controls the supply air flow rate to the thermal zone, and heats up the air using the reheat coils at the box, if required. The control inputs are temperature and flow rate of the air supplied to the zone by its VAV box. The AHU outlet air temperature setpoint $T_a$ is assumed to be constant in this paper. The HVAC system models described in the subsequent paragraphs will follow \cite{kelman2011bilinear,beltran2014optimal,goyal2013occupancy} closely\footnote{Controlling the flow rate is actually more preferable in building codes in consideration of energy efficiency. Herein, we consider both reheat temperature and flow rate are controllable, while the HVAC model with flow rate as the only control input is a simple application of our model.}.

\textbf{State model.} With reference to the notations in Table~\ref{table:control_param}, the continuous time dynamics for the temperature $T^n$ of zone $z_n$ can be expressed as
\begin{align}
\label{eqn:thermal_equ}
C^n\frac{d}{dt}T^n=\mathbf{R}^n\cdot \mathbf{T}+Q^n+\dot{m}_{s}^nc_p(T_{s}^n-T^n)
\end{align}
where the superscript $n$ indicates that the associated quantities are attached to zone $z_n$. $\mathbf{T}\coloneqq [T^1,\cdots,T^N]$ is a vector of temperature of all $N$ zones. $\mathbf{R}^n$ indicates the heat transfer among different zones and outside. $Q^n$ is the thermal load, which can be obtained by applying a thermal coefficient $c_o$ to the number of occupants $V^n$, i.e., $Q^n=c_oV^n$. The control inputs $U^n \coloneqq [\dot{m}_{s}^n,T_{s}^n]$ are the supply air mass flow rate and temperature. Assuming $\dot{m}_{s}^n$, $T_{s}^n$ and $Q^n$ are zero-order held at sample rate $\Delta t$, we can discretize (\ref{eqn:thermal_equ}) using the trapezoidal method and obtain a discrete-time model, which can be expressed as
\begin{align}
\label{eqn:temp_dynamics}
C^n\!\frac{T_{k+1}^n\!-\!T_k^n}{\Delta t}\!\!=\!R^n\!\!\cdot\!\! T_k\!+\!c_o \!V_{k}^n\!+\!\dot{m}_{s,k}^nc_p\bigg(\!\!T_{s,k}^n\!\!-\!\!\frac{T^n_{k+1}\!\!+\!T^n_k}{2}\!\bigg)
\end{align}
where $k$ is the discrete time index and $T_{k}^n = T_t^n|_{t=k\Delta t}$. $Q_{k}^n$, $\dot{m}^n_{s,k}$ and $T_{s,k}^n$ are similarly defined. 

\begin{table}[t!]
\centering
\caption{Parameters used in the HVAC controller.}
\begin{tabular}{c|l|l} \hline
Param. & Meaning & Value \& Units\\\hline
$\Delta t$ & Discretization step & $60s$\\
$c_p$ & Thermal capacity of air &$1 kJ/(kg\cdot K)$\\
$C^n$ & Thermal capacity of the env.& $1000 kJ/K$ \\
$c_o$ & Thermal load per person &$0.1 kW$\\
$R$ & Heat transfer vector & $0 kW/K$\\
$\eta_h$ & Heating efficiency& $0.9$\\
$\eta_c$ & Cooling efficiency &$4$\\
$\beta$ & System parameter &$0.5kW\cdot s/kg$\\
$r_e$ & Electricity price &$1.5\cdot 10^{-4}\$/kJ$\\
$r_h$ & Heating fuel price &$5\cdot 10^{-6}\$/kJ$\\
$\underline{T}$ & Upper bound of comfort zone &$24^\circ C$\\
$\overline{T}$ & Lower bound of comfort zone &$26^\circ C$\\
$T_a$ & AHU outlet air temperature&$12.8^\circ C$\\
$\underline{m_s}$ & Minimum air flow rate &$0.0084kg/s$\\
$\overline{m_s}$ & Maximum air flow rate &$1.5kg/s$\\
$\overline{T}_h$ &Heating coil capacity& $40^\circ C$\\\hline
\end{tabular}
\label{table:control_param}
\end{table}

\textbf{Cost function.} The control objective is to condition the room while
minimizing the energy cost. The power consumption at time $k$ consists of reheating power $P^n_{h,k}= \frac{c_p}{\eta_h}\dot{m}_{s,k}^n(T_{s,k}^n-T_a)$, cooling power $P^n_{c,k}= \frac{c_p}{\eta_c} \dot{m}_{s,k}^n(T_o-T_a)$ and fan power $P^n_{f,k} = \beta\dot{m}_{s,k}^n$,
where $\eta_h$ and $\eta_c$ capture the efficiencies for heating and cooling side, respectively. $\beta$ stands for a system dependent constant. We introduce several parameters to reflect utility pricing, $r_e$ for electricity and $r_h$ for heating fuel. These parameters may vary over time.

Therefore, the total utility cost of zone $z_n$ from time $k=1,\cdots,K$ is $J^n = \sum_{k=1}^K \bigg((r_{e,k}P_{f,k}^n+r_{h,k}P_{h,k}^n + r_{e,k} P_{c,k}^n)\Delta t\bigg)$.

\textbf{Constraints.} The system states and control inputs are subject to the following constraints:

C1: $\underline{T}\leq T_k^n \leq \overline{T}$, comfort range;


C2: $\underline{\dot{m}}_{s}\leq \dot{m}_{s,k}^n\leq \overline{\dot{m}}_{s}$, minimum ventilation requirement and maximum VAV box capacity;

C3: $T_{s,k}^n \geq T_a$, heating coils can only increase temperature;

C4: $T_{s,k}^n \leq \overline{T}_h$, heating coil capacity.

These constraints hold at all times $k$ and all zones $\{z_n\}_{n=1}^N$.

\textbf{MPC controller.} Knitting together the models describ\-ed above, we present an MPC-based control strategy for the HVAC system to efficiently accommodate for occupancy variations. In this control algorithm, we assume that 
the predicted occupancy during the optimization horizon to be the same as the instanteneous occupancy observed at the beginning of control horizon. It was shown to be in~\cite{goyal2013occupancy} that the control algorithm with this assumption can achieve comparable performance with the MPC that constructs explicit occupancy model to predict occupancy for future time steps.

Let $U_{1:K}^{1:N}$ be the shorthand for $\{U_k^n|k=1,\cdots,K,\quad n= 1,\cdots,N\}$. The optimal control inputs for the next $K$ time steps are obtained by solving $\min_{\substack{U_{1:K}^{1:N}}} \sum_{n=1}^NJ^n$,
subject to the inequality constraints C1-C4 and the equality constraint (\ref{eqn:temp_dynamics}) and $T_1^n = T_{init}^n$, $\forall n = 1,\cdots,N$, where $T_{init}^n$ is the initial temperature of zone $z_n$ at each MPC iteration. We can see that the optimal control input is a function of the distorted occupancy that the controller sees and the initial temperature. We express this relationship explicitly by denoting the optimal control action at zone $z_n$ as $U_{MPC}^n(V^n,T_{init}^n)$ . In addition, the energy cost incurred by applying the optimal control action is denoted by $J_{MPC}^n(U_{MPC}^n(V^n,T_{init}^n),Y^n)$, where the second argument stresses that the actual control cost is dependent on the real occupancy.

\section{Privacy-Enhanced Control}
\label{sec:framework}

With the HVAC model established, we can now develop the mathematical framework to discuss a privacy-enhanced architecture. We will first introduce MI as the metric we use throughout the paper to quantify privacy, and then present a method to optimally design the distortion mechanism which minimizes the privacy loss within a pre-specified constraint on control performance.

%
\subsection{Privacy metric}
\begin{definition}\cite{cover2012elements}
For random variables $X$ and $V$, the \emph{mutual information} is given by:
\begin{align}
I(X;V) = H(X) - H(X \vert V)
\end{align}
where $H(X)$ and $H(X\vert V)$ represent \emph{entropy} and \emph{conditional entropy}, respectively. Let $\Pbb_X(x) = \Pbb(X=x)$, $H(X)$ and $H(X\vert V)$ are defined as
\begin{align}
&H(X) = -\sum_{x} \Pbb_X(x) \log( \Pbb_X(x) )\\
&H(X \vert V)\! = \!- \!\sum_{v}\Pbb_V(v)\! \bigg( \!\!\sum_{x}\Pbb_{X|V}\!(x \vert v) \log\!\big(\Pbb_{X|V}\!(x \vert v)\big) \!\bigg)
\end{align}
\end{definition}
\textbf{Remark.} Entropy measures uncertainty about $X$, and conditional entropy can be interpreted as the uncertainty about $X$ after observing $V$. By the definition above, MI is a measure of the reduction in uncertainty about $X$ given knowledge of $V$. We can see that it is a natural measure of privacy since it characterizes how much information one variable tells about 
another. It is also worth noting that inference technologies evolve and MI as a privacy metric does not depend on any particular adversarial inference algorithm~\cite{RajagopalanSankarMohajerEtAl2011} as it models the statistical relationship between two variables.

In this paper, we will be using the MI between location traces and occupancy observations, i.e., $I(X_k^{(1:M)};V_k^{1:N})$, as a metric of privacy loss. This metric reflects the reduction in uncertainty about location traces $X_k^{(1:M)}$ due to observations of $V_k^{1:N}$. As a proof of concept, we will verify that this metric serves as an accurate proxy for an adversary's ability to infer individual location traces in the experiments. We further introduce some assumptions which allow us to simplify the expression of the privacy loss and obtain a form of MI that has direct relationship with the distortion mechanism $P(V_k^n|Y_k^n)$ we wish to design.

Based on results in ergodic theory~\cite{Kallenberg2002}, we know that the probability distribution of individual location traces will converge to a unique stationary distribution under very mild assumptions\footnote{Since there are only finitely many zones, a sufficient condition is the existence of a path from $z_i$ to $z_j$ with positive probability for any two zones $z_i$ and $z_j$.}. For more details on stationary distributions, we refer the readesr to~\cite{Kallenberg2002}. This observation justifies the following:

\begin{assumption}
\label{ass:stationary}
The Markov chains $X_k^{(m)}$ have a unique stationary distribution for all occupants $o_m$ and are distributed according to those stationary distributions for all time steps $k$.
\end{assumption}

Combining this assumption and the occupancy-location model we presented in the preceding section, we present a proposition that allows us to great simplify the form of the privacy loss:
\begin{proposition}
\label{prop:reduce}
By 
Assumption~\ref{ass:loc_cond_indep}, we have that:
\begin{align}
I(X_k^{(1:M)}; V_k^{1:N}) = I(Y_k^{1:N}; V_k^{1:N})
\end{align}

By Assumption~\ref{ass:stationary}, we have that $I(Y_k^{1:N}; V_k^{1:N})$ is a constant for all $k$, so we will drop the subscript: $I(Y^{1:N}; V^{1:N})$.

Finally, by the various conditional independences introduced in Assumption~\ref{ass:loc_cond_indep}: \begin{align}
I(Y^{1:N}; V^{1:N}) =\sum_{n = 1}^N I(Y^n; V^n)
\end{align}
\end{proposition}
\textbf{Remark.} The result that $I(Y_k^{1:N}; V_k^{1:N})$ is a constant value for all $k$ allows us to design a single distortion mechanism $P(V^n|Y^n)$ for all time steps (note that we drop the subscript $k$ to indicate the time-homogeneity of the distortion mechanism). By Proposition~\ref{prop:reduce}, minimization of privacy loss $I(X_k^{(1:M)}; V_k^{1:N})$ can be conducted by minimizing a simpler expression $\sum_{n = 1}^N I(Y^n; V^n)$.


\subsection{Optimal distortion design}
We wish to find a distortion mechanism $P(Y^n|V^n)$ that can produce some perturbed occupancy data with minimum information leakage, while the performance of the controller using the perturbed occupancy data is on a par with that using true occupancy. To be specific, we will bound the difference of energy costs incurred by the controllers seeing distorted and real occupancy data.

Let $T_{init1}$ and $T_{init2}$ be initial temperature of the controller using distorted and real occupancy, respectively. Recall that $U_{MPC}^n(V^n,T_{init}^n)$ and $J_{MPC}^n(U_{MPC}^n(V^n\!\!,$ $T_{init}^n),Y^n)$ stand for the optimal control actions and the associated cost based on the distorted occupancy; correspondingly, if the controller sees the real occupancy data, the optimal control action and the associated cost will be $U_{MPC}^n(Y^n,T_{init}^n)$ and $J_{MPC}^n(U_{MPC}^n(Y^n,T_{init}^n),Y^n)$, respectively. We denote the resulting temperature after applying optimal control actions as $T_{MPC}^n(U_{MPC}^n(V^n,T_{init}^n),Y^n)$, where the second argument emphasizes that the temperature evoluation depends on the true occupancy. We introduce the following constraints: 
$\forall |T_{init1}-T_{init2}|\leq \Delta_T'$, $y=0,\cdots,M$, $n=1,\cdots,N$,

\textbf{C5: Cost difference constraint}
\begin{align}
\label{eqn:opt_cost}
&E_{\mathbb{P}(V^n \vert Y^n=y)}\bigg[ J_{MPC}^n\big(U^n_{MPC}(T_{init1},V^n),y\big) - \nonumber\\  
&\quad\quad\quad\quad J_{MPC}^n\big(U_{MPC}^n(T_{init2},y),y\big)\bigg]\leq \Delta
\end{align}

\textbf{C6: Resulting temperature constraint}
\begin{align}
\label{eqn:opt_init_temp}
&E_{\mathbb{P}(V^n \vert Y^n=y)}\bigg[\big|T_{MPC}^n\big(U_{MPC}^n(T_{init1},V^n),y\big)-\nonumber\\
&\quad\quad\quad \quad T_{MPC}^n\big(U_{MPC}^n(T_{init2},y),y\big)\big|\bigg]\leq \Delta_T
\end{align}

C5 states that the cost difference between using the distorted occupancy measurements $V^n$ and using the ground truth occupancy measurements $Y^n$ is bounded by $\Delta$ in expectation, for any possible value of $Y^n$. The cost difference can be regarded as the control performance loss due to the usage of distorted data, and $\Delta$ stands for the tolerance on the control performance loss. C5 alone is a one-step performance guarantee, that is, it only bounds the cost difference associated with a single MPC iteration. In practice, MPC is repeatedly solved from the new initial temperature, yielding new control actions and temperature trajectories. In order to offer a guarantee for future cost difference, we introduce another constraint C6 on the resulting temperature difference of one MPC iteration. The idea is that the resulting temperature will become the new initial temperature of the next MPC iteration. If the resulting temperature difference between using distorted occupancy data and using true occupancy data is bounded within a small interval $\Delta_T$, in the next MPC iteration C5 will provide a bound on cost difference for new initial temperatures that do not differ too much, since the cost difference constraint C5 is imposed to hold for all $|T_{init1}-T_{init2}|\leq \Delta_T'$. Typically, $\Delta_T'$ is set to be similar to $\Delta_T$, but a small value of $\Delta_T'$ is preferred in order to assure the feasibility of the optimization problem (since the number of constraints increases with $\Delta_T'$).

Now, we are ready to present the main optimization for privacy-enhanced HVAC controller by combining the privacy metric and performance constraint just presented. Suppose the assumptions of Proposition~\ref{prop:reduce} hold. Given the control performance loss tolerance $\Delta$, the \emph{optimal distortion mechanism} is given by solving:
\begin{align}
\label{eqn:main_opt}
\min_{\substack{\mathbb{P}(V^n \vert Y^n)\\n=1,\cdots,N}} ~& \sum_{n = 1}^N I(Y^n; V^n)
\end{align}
subject to the constraint C5-C6. $\Delta$ serves as a knob to adjust the balance between privacy and the controller performance loss. Increasing $\Delta$ leads to larger feasible set for the optimization problem, and thus a smaller value of MI (or privacy loss) is expected. Using the methodology presented in Section~\ref{sec:model}, we are able to calculate 
the terms inside the expectation in (\ref{eqn:opt_cost}) and (\ref{eqn:opt_init_temp}) for all $|T_{init1}-T_{init2}|\leq \Delta_T'$ and $y=0,\cdots,M$. Treating these as constants, calculating the optimal privacy-aware sensing mechanism is a convex optimization program, and can be efficiently solved. Additionally, since the constraints are enforced for each zone, the optimization (\ref{eqn:main_opt}) can actually be decomposed to $N$ sub-problems and thus we can solve the optimal distortion scheme separately for each zone.

\textbf{Remark on noisy occupancy data.} In the preceding privacy-enhanced framework, we consider the occupancy can be accurately detected. In practice, the occupancy data may be noisy itself, and thereby the distortion mechanism will be designed based on noisy occupancy $W_k^n$ instead of true occupancy $Y_k^n$. In effect, the distortion designed using noisy occupancy provides an upper bound on the privacy loss. That is, in practice we could use noisy occupancy to design the distortion mechanism and the realized privacy loss can only be lower than the minimum privacy loss obtained from the optimization. Note that we have the Markov relationship: $Y_k^n\rightarrow W_k^n\rightarrow V_k^n$ when the distortion is applied to noisy data. Then the proof follows from the data processing inequality~\cite{cover2012elements}.


\section{Evaluation}
\label{sec:eval}


\subsection{Experiment Setup}
\textbf{Occupancy dataset.} The occupancy data used in this paper is from the Augsburg Indoor Location Tracking Benchmark~\cite{ea142004augsburg}, which includes location traces for $4$ users in a office building with $15$ zones. The location data in the benchmark dataset was recorded every second over a period of $4$ to $9$ weeks. Since the dataset contains some missing observations due to technical issues or the vacation interruption, we finally use the dataset from November 5th to 24th in our experiment, during which the location traces of all the $4$ users are complete, and subsample the dataset with $1$-minute resolution. The ground truth occupancy data was synthesized by aggregating the locations trace of each user. Table~\ref{table:data_stat} shows two statistics of the benchmark dataset. Notably, of all transitions per day, $66.7\%$ to $84.6\%$  either start from or end at one's own office, and office location can divulge one's identity. This sheds light on why location traces of individual users can be actually inferred from the ``anonymized" occupancy data. 

\begin{table}[ht]
\centering
\caption{The average number of transitions each user made in each workday, and the average percentage of transitions from or to one's office.}
\begin{tabular}{c|c|c} \hline
User&avg \# of transitions & avg \% of transitions\\
&per day& from/to office per day\\\hline
1 & 9.3 & 84.6\%\\\hline
2 & 20.2 & 75.4\%\\\hline
3& 9.9 & 66.7\%\\\hline
4& 7.6 &75.5\%\\\hline
\end{tabular}
\label{table:data_stat}
\end{table}

\textbf{Adversary inference.} We consider the adversary to be an \textit{insider} with authorized building automation system access. One can think of it as the worst case of privacy breach, because insiders not only learn the ancillary information that is public-available, but are familiar with building operation policies. To be specific, the following auxiliary information is assumed to be available to the adversary: (1) Building directory and occupant mobility patterns, encoded by the transition matrix of each occupant\footnote{
In the experiment, we use $4$ days' occupancy data and $2$ days' location traces to learn these parameters and the rest for evaluating our framework. 
}; (2) Occupancy distortion mechanism designed by building manager.

The adversary attempts to reconstruct the most probable location trace given the occupancy data and the auxiliary information. That is, the attack is to find the MAP of location traces given the other information. The approach to finding MAP is well known as Viterbi algorithm in HMM. However, Viterbi is infeasible in the FHMM case as the location traces to be solved reside in a exponentially large state space ($N^M\times K$). We propose a fast inference method based on Mixed Integer Programming, and thus more efficiently evaluate the adversary's inference attack. The interested readers are referred to the code implementation of this paper for the details of the fast inference algorithm. 

\textbf{Controller parameters.} Without loss of generality, we consider the zones have the same thermal properties. The comfort range of temperature in the zones is defined to be within $24-26^\circ C$ as in~\cite{nagarathinam2015centralized}. The minimum flow rate is set to be $0.084 kg/s$ to fulfill the minimum ventilation requirement for $25 m^2$-sized zone as per ASHRAE ventilation standard 62.1-2013~\cite{american2013}. The optimization horizon of the MPC is $120$ min, and the control commands are solved for and updated every $15$ min~\cite{goyal2013occupancy}. Other design parameters are shown in Table~\ref{table:control_param}, which bascially follows the choices in~\cite{kelman2011bilinear}. 

\textbf{Platform.} The algorithms are implemented in MATLAB; The interior-point algorithm is used to solve the bilinear optimization problem in MPC. To encourage the research on the privacy-preserving controller, the codes involved in this paper will be open-sourced in \url{http://people.eecs.berkeley.edu/~ruoxijia/code}.

\subsection{Results}
\subsubsection{MI as proxy for privacy}
\begin{figure}[ht]
\centering
\includegraphics[width=0.8\columnwidth,trim={0.2cm 0.1cm 0.2cm 0cm},clip]{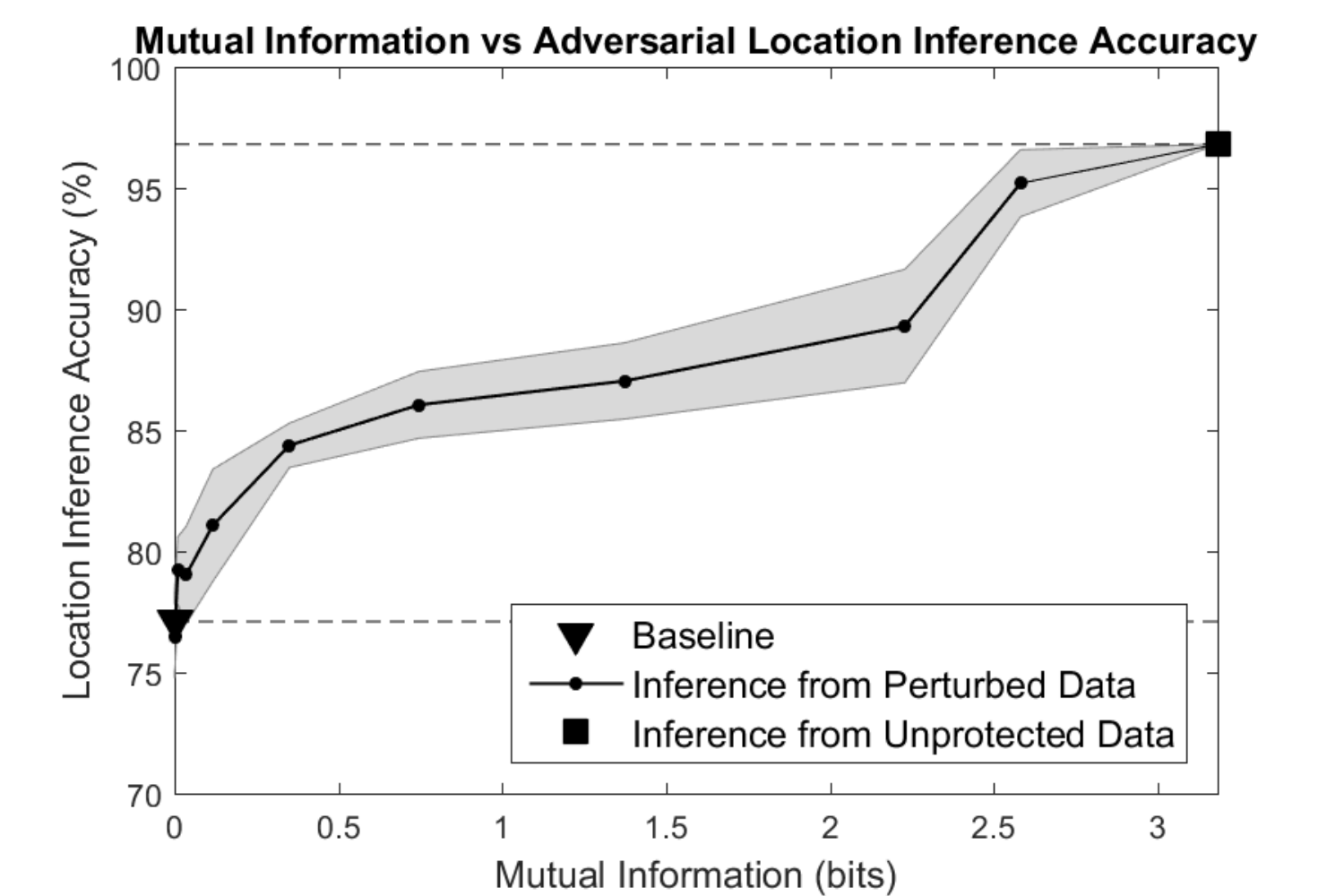}
\caption{The adversary location inference accuracy increases as MI increases. The black line and the band around it show the mean and standard deviation of inference accuracy across ten MC simulations, respectively. The black square shows the location inference accuracy if the adversary sees true occupancy data. The black triangle gives the accuracy when the adversary outputs a constant location estimate.}
\label{fig:mi_vs_acc}
\end{figure}

We solve the MI optimization for different tolerance levels of control performance deterioration due to the usage of the distorted data, i.e., $\Delta$, and obtain a set of optimal distortion designs and corresponding optimal values of MI. We then randomly perturb the true occupancy data using the different distortion designs, and infer location traces from the perturbed occupancy data. Monte Carlo (MC) simulations are carried out to assess results under the random distortion design. The inference accuracy is defined to be the ratio between the counts of correct location predictions over the total time steps. Figure~\ref{fig:mi_vs_acc} demonstrates the monotonically increasing relationship between adversarial location inference accuracy and MI, which justifies the usage of MI as a measure of privacy loss. When the adversary has perfect occupancy data, individual location traces can be inferred with accuracy of $96.81\%$. On the contrary, when the MI approaches zero, the adversary tends to estimate the location of each user to be constantly outside of the building, which is the best estimate the adversary can generate based on the uninformative occupancy data since people spend most of their time in a day outside. In this case, the inference accuracy is $77\%$ but the adversary actually has no knowledge about users' movement. This serves as a baseline of the adversarial location inference performance. 

\subsubsection{Utility-Privacy Trade-off}
\begin{figure}[!t]
\centering
\includegraphics[width=0.8\columnwidth,trim={0.2cm 0.1cm 0.2cm 0cm},clip]{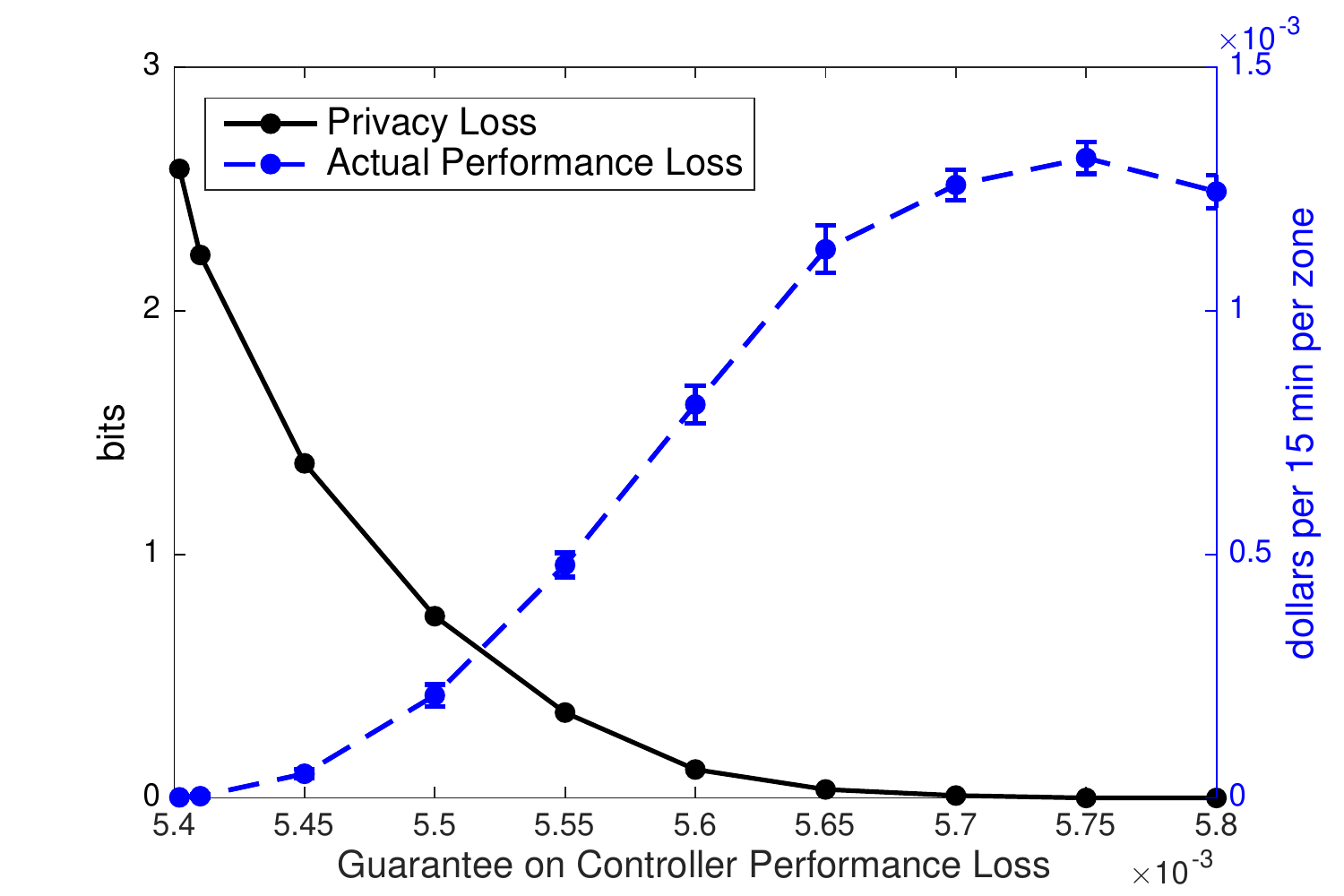}
\caption{The changes of MI and actual control cost difference between using true and perturbed occupancy as the theoretical control cost difference changes. The blue dot line and errorbar demonstrate the mean and standard deviation of actual control cost difference across ten MC simulations, respectively.}
\label{fig:trade_off}
\end{figure}

Figure~\ref{fig:trade_off} shows the variation of privacy loss and controller performance loss with respect to different choices of $\Delta$, which is the theoretical guarantee on controller performance loss. It is evident that privacy loss and control performance loss exhibit opposite trends as $\Delta$ changes. The privacy loss, measured by MI, monotonically decreases as $\Delta$ gets larger. This is the manifestation of the intrinsic utility-privacy trade-off embedded in the main optimization problem (\ref{eqn:main_opt}). As the performance constraint $\Delta$ is more relaxed, a smaller value of MI can be attained and thus privacy can be better preserved. The actual performance loss, measured by the HVAC control cost difference (between using distorted and true data) averaged across different MPC iterations and difference zones, generally increases with $\Delta$ and is upper bounded by $\Delta$. This indicates that the theoretical constraint on controller performance loss in our framework is effective and can actually provide a guarantee on the actual controller performance. We can see that the bound is far from tight, since the framework enforces the constraints on the controller performance for every possible true occupancy value to ensure the robustness while in practice the occupancy distribution is very spiked about the mean occupancy.

\begin{figure}[!t]
\centering
\includegraphics[width=0.8\columnwidth,trim={0.2cm 0cm 0.2cm 0cm},clip]{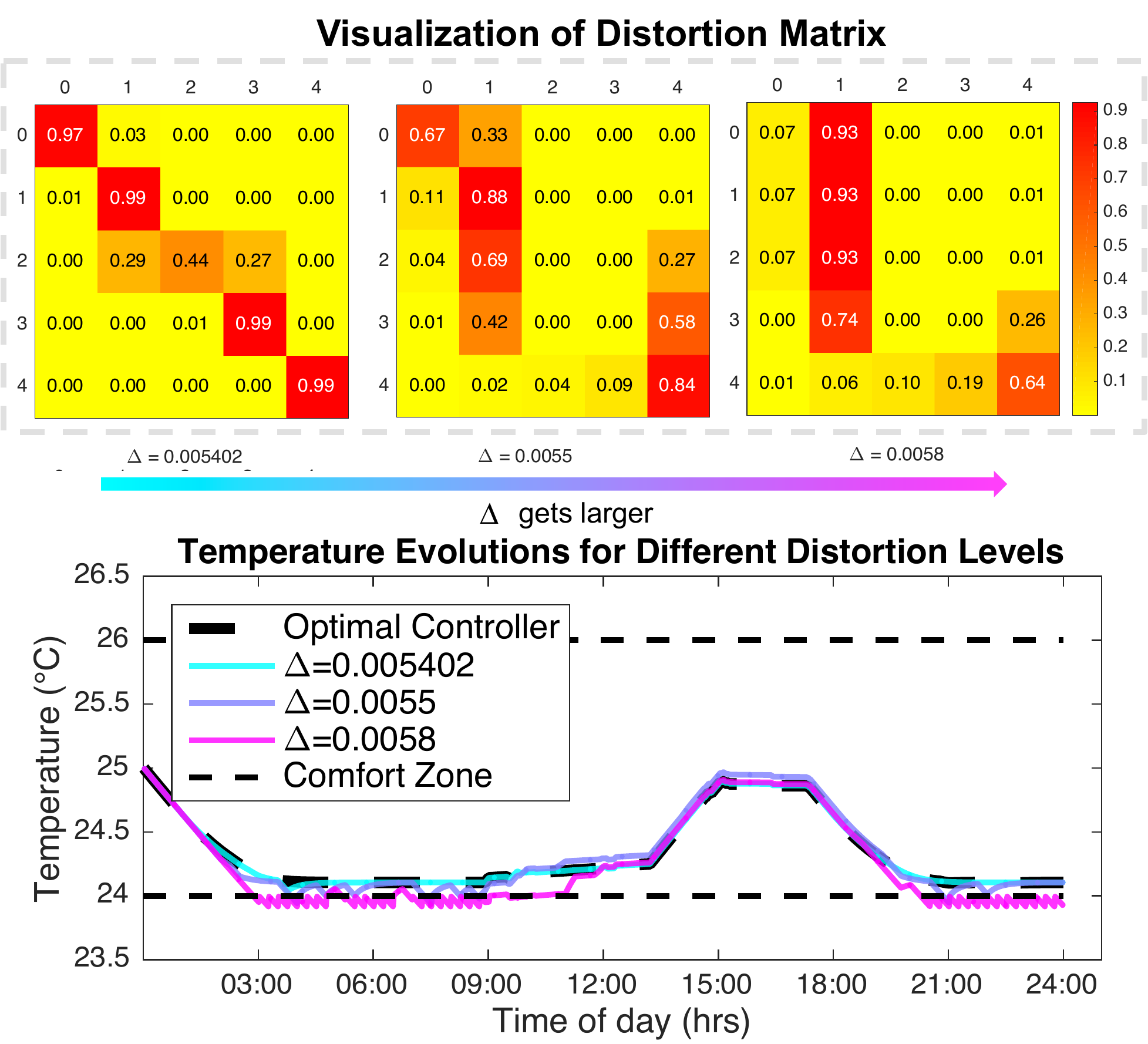}
\caption{Illustration of distortion matrix $P(V|Y)$ under different controller performance guarantees. The row index corresponds to the value of $Y$, while colomn index corresponds to $V$. The zone temperature traces resulted from the controllers using occupancy data that is randomly distorted by different distortion matrices are also shown.}
\label{fig:dist_mat}
\end{figure}

Figure~\ref{fig:dist_mat} visualizes the distortion mechanism obtained by solving the MI under different choices of the tolerance on the control performance loss $\Delta$. It can be clearly seen that the mechanism creates a higher level of distortion as $\Delta$ increases. When $\Delta$ is small, the resulting distortion matrix assigns most probability mass on the diagonal, i.e., the occupancy is very likely to keep unperturbed. As $\Delta$ gets larger, the distortion mechanism tends to have the same rows, in which case the distribution of distorted occupancy data is invariant under the change of true occupancy and MI between true occupancy and perturbed occupancy, i.e., the privacy loss, tends to be zero. We also plot the temperature evolution under different distortion levels. Since we enforce a hard constraint on temperature, we can see that the zone temperature stays within the comfort zone for all $\Delta$'s. However, larger $\Delta$ would lead to a larger deviation from the temperature controlled using the true occupancy. 

\subsubsection{Comparison with Other Methods}
\begin{figure}[t]
\centering
   \subfloat[\label{fig:comp_scheme}]{%
    \includegraphics[width=\columnwidth,trim={0cm 0.1cm 0cm 0.1cm},clip]{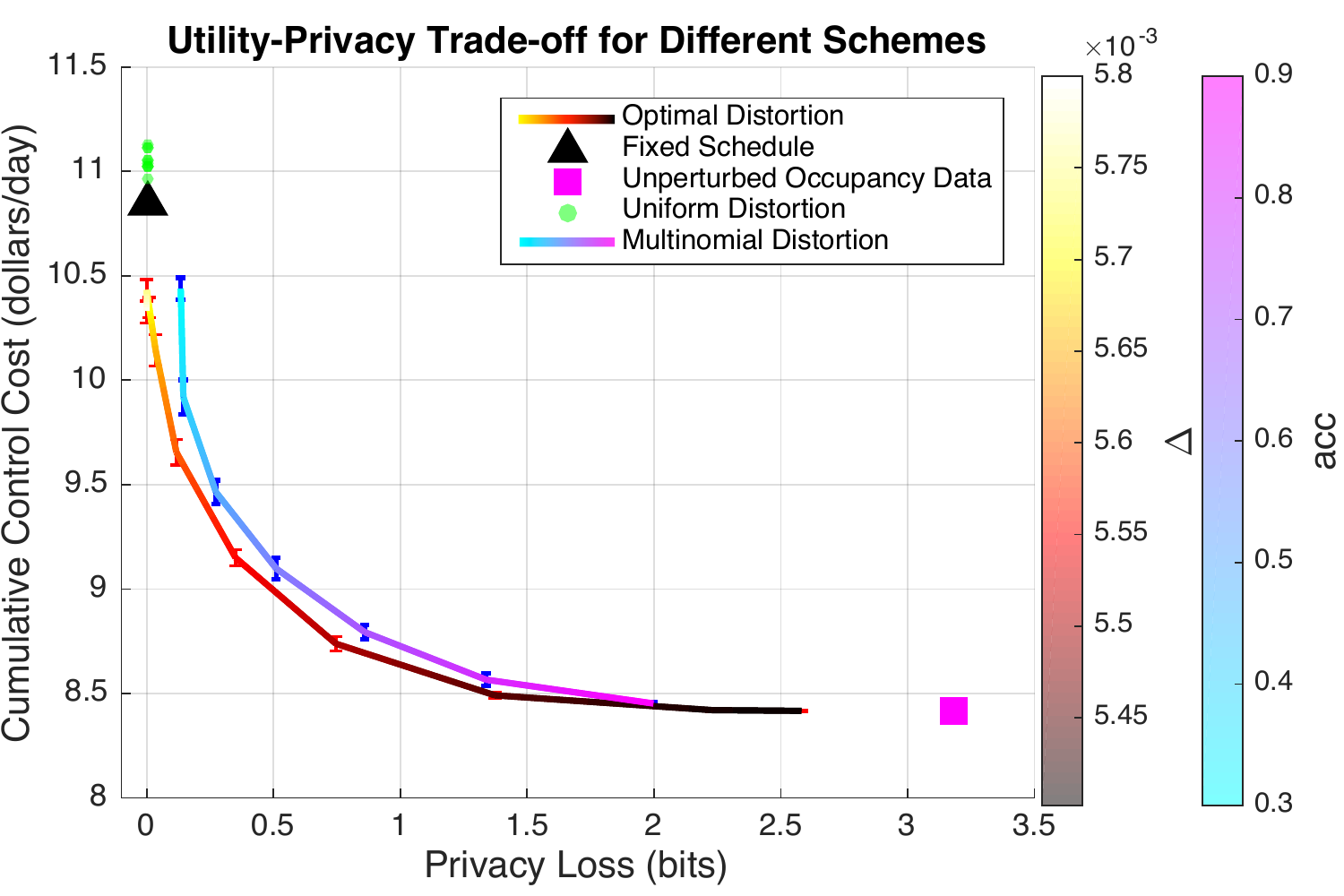}    
   }\\
    \subfloat[\label{fig:comp_scheme_sim}]{%
    \includegraphics[width=\columnwidth,trim={0cm 0.1cm 0cm 0.1cm},clip]{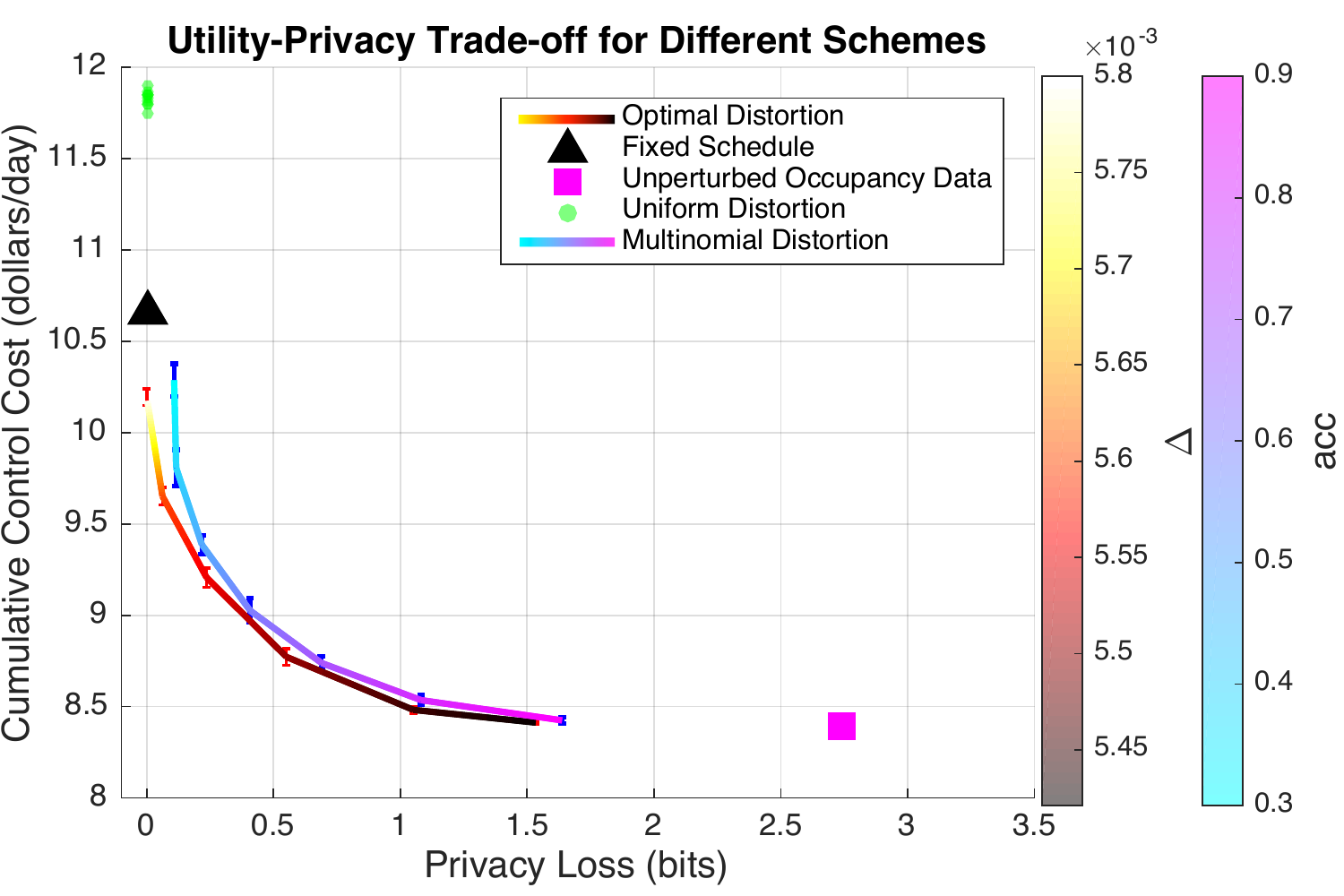}
    }
    \caption{Comparison of the privacy-utility trade-off of controllers using different forms of occupancy data, evaluated based on (a) real-world occupancy data and (b) synthesized data.}
  \end{figure}

We compare the performance of the HVAC controller using our optimally perturbed data against using unperturbed occupancy data, fixed occupancy schedule as well as randomly perturbed data by other distortion methods. In Figure~\ref{fig:comp_scheme} we plot the privacy loss and control cost for controllers that use the various forms of occupancy data. Fixed occupancy schedule (assuming maximum occupancy during working hours and zero otherwise) exposes zero information about individual location traces, but cannot adapt to occupancy variations and thus incurs considerable control cost. The controller based on clean occupancy data is most cost-effective but discloses maximum private information. One of the random distortion method to be compared is uniform distortion scheme in which the true occupancy is perturbed to some value between zero to maximum occupancy with equal probability. We carry out $10$ MC simulations to obtain the control cost incurred under this random perturbation scheme. It can be seen that the uniform distortion scheme protects the private information with compromised controller performance. 

A natural question arising is if the current occupancy sensing systems provide intrinsic privacy-preserving features as there always exists occupancy estimation errors. Can we use a cheaper and inaccurate occupancy sensor to acqiure privacy? As is suggested by the occupancy sensing results in~\cite{jin2015sensing}, the estimation noise of a real occupancy sensing system can be modeled by a multinomial distribution which has most probability mass at zero. Inspired by this, we use the following multinomial distortion schemes to imitate a real occupancy sensing system with disparate accuracies $acc$,
\begin{align}
P(V^n|Y^n=y) = \left\{
\begin{array}{ll}
acc, \quad V=y\\
\frac{1-acc}{2},V\!=\!y\!-\!1 \text{ or } y\!+\!1 \text{ if } y\!\neq\! 0\\
\frac{1-acc}{2},\quad V \!= \!1 \text{ or } 2, \text{ if } y\!=\!0
\end{array}
\right.
\end{align}
Again, MC simulations are performed to evaluate the control performance under this random perturbation, and the results are shown in Figure~\ref{fig:comp_scheme}. It can be seen that when the privacy loss is relatively large (or data is slightly distorted), the control cost of our optimal noising scheme and the multinomial noising scheme do not differ too much. This is because at this level of privacy loss the two distortion schemes behave similarly, as shown in Figure~\ref{fig:dist_mat}, where the occupancy keeps untainted with high probability. But as the privacy loss decreases, our optimal noising scheme's intelligent noise placement begins to significantly improve control performance. In addition, our optimal distortion Pareto dominates the other schemes.

To investigate the scalability of our proposed scheme, we create synthetic data that simulates location traces for $15$ occupants based on the Augsburg dataset. We extract the occupants' movement profile, i.e., transition parameters, from the original dataset and randomly assign the profiles to synthesized occupants. An occupant randomly chooses the next location according to the movement profile. The privacy-utility curve evaluated on this larger synthesized dataset is illustrated in Figure~\ref{fig:comp_scheme_sim}, which demonstrates that the optimality of our distortion scheme is preserved when the experiment is scaled up. We can see that the privacy loss of the controller using the unperturbed occupancy gets lower when incorporating more occupants. Although privacy risks are lower as we scale up the experiment since with more people sharing the space it will be more difficult to identify each individuals, adding distortion to occupancy measurements can preserve the privacy even further as shown in Figure~\ref{fig:comp_scheme_sim}.

\section{Conclusions}
In this paper, we present a tractable framework to model the trade-off between privacy and controller performance in a holistic manner. We take occupancy-based HVAC controller as an example where the objective is to utilize occupancy data to enable smart controls over the HVAC system while protect individual location information from being inferred from the occupancy data. We use MI as the measure of privacy loss, and formulate the privacy-utility trade-off by a convex optimization problem that minimizes the privacy loss subject to a pre-specified controller performance constraint. By solving the optimization problem, we can obtain a mechanism that injects optimal amount of noise to occupancy data to enhance privacy with control performance guarantee. We verify our framework using real-world occupancy data and simulated building dynamics. It is shown that our theoretical framework is able to provide guidelines for practical privacy-enhanced occupancy-based HVAC system design, and reaches a better balance of privacy and control performance compared with other occupancy-based controllers.

\label{sec:conclusion}


%
\bibliographystyle{abbrv}
\bibliography{mybib,DONG_ROY_refs}  
%
%



\end{document}